\documentclass{article}

\usepackage{amsfonts,mathtools,bm,mathrsfs}
\usepackage[OT1]{fontenc}
\usepackage[inline]{enumitem}
\usepackage{siunitx}
\usepackage[dvipsnames]{xcolor}
\RequirePackage{amsthm}
\usepackage[top=1.25in, bottom=1.25in, left=1.25in, right=1.25in]{geometry}

\theoremstyle{definition}
\newtheorem{definition}{Definition}
\newtheorem{example}{Example}
\newtheorem{remark}{Remark}

\def\eqns#1{\begin{equation*}#1\end{equation*}}
\def\eqnl#1#2{\begin{equation}\label{#1}#2\end{equation}}
\def\eqnsa#1{\begin{align*}#1\end{align*}}

\def\one{\mathbf{1}}

\def\sfE{\mathsf{E}}

\def\sfX{\mathsf{X}}

\def\sfZ{\mathsf{Z}}

\def\calE{\mathcal{E}}

\def\calV{\mathcal{V}}

\def\calX{\mathcal{X}}
\def\calY{\mathcal{Y}}
\def\calZ{\mathcal{Z}}

\def\bbE{\mathbb{E}}

\def\bbP{\mathbb{P}}
\def\bbR{\mathbb{R}}

\def\bbV{\mathbb{V}}

\def\uvs{\bm{s}}

\def\uvy{\bm{y}}

\def\uvtheta{\bm{\theta}}

\def\uvpsi{\bm{\psi}}

\def\c{\mathrm{c}}
\def\d{\mathrm{d}}

\def\N{\mathrm{N}}

\def\ind#1{\one_{#1}}

\def\given{\,|\,}
\def\param{;}
\def\Given{\,\big|\,}
\def\AND{\qquad\text{and}\qquad}
\def\et{,\;}

\def\st{:}

\DeclareMathOperator*{\argmax}{argmax}

\DeclareMathOperator{\Bel}{Bel}
\DeclareMathOperator{\Pl}{Pl}

\def\r{\mathrm{r}}
\def\u{\mathrm{u}}

\def\oP{\bar{P}}

\def\bscdot{\bm{\cdot}}
\def\dotparam{\bscdot\,\param}

\begin{document}


\title{Parameter estimation with a class \\ of outer probability measures}
\author{Jeremie Houssineau\\
Department of Statistics\\
University of Warwick}
\date{}
\maketitle

\begin{abstract}
We explore the interplay between random and deterministic phenomena using a representation of uncertainty based on the measure-theoretic concept of outer measure. The meaning of the analogues of different probabilistic concepts is investigated and examples of application are given. The novelty of this article lies mainly in the suitability of the tools introduced for jointly representing random and deterministic uncertainty. These tools are shown to yield intuitive results in simple situations and to generalise easily to more complex cases. Connections with Dempster-Shafer theory, the empirical Bayes methods and generalised Bayesian inference are also highlighted.
\end{abstract}

\noindent%
{\it Keywords:} Deterministic uncertainty, Outer measure, Bayesian inference

\section{Introduction}

Alternatives to the standard probabilistic approach have been advocated by eminent scientists throughout the 20th century \cite{Fisher1930, Dempster1967, Zadeh1978}. These ideas are common in some areas, such as expert systems \cite{Walley1996} and statistical signal processing \cite{Basir2007}, and continue to be studied and generalised \cite{Hannig2009, Martin2010}; yet, they have not widely been adopted by the statistical community. In parallel, the ever-growing complexity of statistical models is becoming a challenge for standard Bayesian inference by requiring increasing degrees of flexibility and robustness, and fields related to statistics such as machine learning are in need of scalable means of uncertainty quantification.

Possibility theory \cite{Shackle1961, Zadeh1978, Dubois2015} provides a representation of deterministic uncertainty, i.e.\ uncertainty due to a lack of information, based on optimisation rather than integration. This could be instrumental in devising scalable and principled methods where some parameters are set to their most-likely value rather than integrated out. The tools of possibility theory also allow for modelling a complete absence of information about a quantity of interest which could address existing challenges in modelling prior information about (hyper-)parameters of statistical models.

The objective in this article is to show that possibility theory and probability theory can be rigorously combined so as to provide a general framework for representing statistical quantities in which random and deterministic uncertainties are intertwined. The suggested way of augmenting the probabilistic paradigm with a dedicated representation of deterministic uncertainty relies on the standard measure-theoretic notion of outer measure \cite{Caratheodory1909}. Although the use of this notion is often limited to foundational measure theory, we show that a specific class of outer measures can be used for statistical inference in a way that is similar to the standard probabilistic approach.

\subsection*{Notations}

The theoretical foundations of standard statistical inference have been thoroughly studied over the 20\textsuperscript{th} century so that the subject can now be discussed in a more pragmatic way. However, this article introduces original objects and bypassing the basic verifications could have serious consequences. Some measure-theoretic remarks will therefore be necessary along the way; yet, these remarks can often be ignored without altering the accessibility of the results. Most of the concepts in this article are described using two sets, $\Theta$ and $\sfX$, the latter being assumed to have all the required properties for defining probability distributions on it. When defining a probability distribution $p$ on $\sfX$, it will often be convenient to deal directly with the integral of a function $\varphi$ on $\sfX$ against $p$, which we will denote by
$$
p(\varphi) = \int \varphi(x) p(x) \d x,
$$
where the same notation has been used for the probability distribution $p$ and its probability density function (p.d.f.). We also write $p(B) = \int_B p(x) \d x = p(\ind{B})$ for any subset $B$ of $\sfX$, with $\ind{B}$ the indicator function of $B$, i.e.\ $\ind{B}(x)$ equals to $1$ if $x \in B$ and to $0$ otherwise. If $f$ is a function on $\Theta$ then
$$
\sup f = \sup_{\theta \in \Theta} f(\theta) = \sup \{ f(\theta) : \theta \in \Theta\}
$$
is the supremum of $f$ over $\Theta$. These three notations will be used interchangeably. Note that the term function will exclusively be used for real-valued mappings.

\begin{remark}
Although p.d.f.s are used throughout the article, most of the results can be stated with probability measures on arbitrary Polish spaces. All the considered subsets and mappings of measurable spaces will be implicitly assumed to be measurable.
\end{remark}

\subsection*{Approach and relation to other work}

The proposed approach starts from the same premise as possibility theory \cite{Shackle1961, Dubois2015} and Dempster-Shafer theory \cite{Dempster1967,Shafer1976} and considers that the additivity assumption of probability distributions must be relaxed in order to model a lack of knowledge. If $\theta^*$ is a fixed but unknown parameter in a set $\Theta$ then we might not have enough information to define a probability distribution $p$ modelling the uncertainty in $\theta^*$ in such a way that, e.g., $p(B) + p(B^{\c}) = 1$ for any $B \subseteq \Theta$, with $B^{\c}$ the complement of $B$ in $\Theta$. Defining $p(B)$ as being proportional to the size of $B$ is often the least informative choice, but this approach leads to issues when $\Theta$ is unbounded since $p(\Theta) = \infty$ in this case. Instead, if the assumption of additivity is removed, then one can simply set the \emph{credibilities} of the ``events'' $\theta^* \in B$ and $\theta^* \in B^{\mathrm{c}}$ to $1$ in order to express that there is no objection against either of these events. A set function which has all the properties of a probability distribution except for additivity is an outer measure, say $\bar{P}$, which verifies $\bar{P}(\Theta) = 1$. We will therefore refer to such set functions as \emph{outer probability measures} (o.p.m.s).

In order to formalise the usual notions associated with random variables to this generalised context, we will introduce in Section~\ref{sec:uncertainVariable} the notion of \emph{uncertain variable}. Informally, we use an uncertain variable $\uvtheta$ to represent the different possible values that the fixed parameter $\theta^*$ can take. In this situation, the information we hold about $\theta^*$ will modelled by an o.p.m.\ that is nowhere additive. A simple way to define such an o.p.m.\ is as
$$
\bar{P}_{\uvtheta}(A) = \sup_{\theta \in A} f_{\uvtheta}(\theta), \qquad A \subseteq \Theta,
$$
where, by construction, $f_{\uvtheta}$ is a non-negative function with supremum equal to $1$. We will refer to this type of function as \emph{possibility functions} although they are called possibility distributions in the context of possibility theory. This is to better highlight the difference between these and probability distributions. Possibility functions can be used in a way that is strikingly similar to probability distributions, except that integrals will be replaced by supremums and sums by maximums. Some of their properties will be covered in Section~\ref{sec:uncertainVariable}.

The objective of this article is tackle the case where a quantity of interest might be under the influence of both random and deterministic but unknown phenomena. This is best illustrated with an important class of o.p.m.\ corresponding to one of the most fundamental object in statistical inference: a parametric family of probability distributions. We consider a parameter of interest in a set $\Theta$ and the uncertainty caused by the randomness associated with each parameter value $\theta \in \Theta$ via a random variable on $\sfX$ with law $p(\dotparam \theta)$, parametrised by $\theta \in \Theta$. The uncertainty in the true parameter value in $\Theta$ can be modelled as before by an uncertain variable $\uvtheta$ described by a possibility function $f_{\uvtheta}$ on $\Theta$. The associated quantity in $\sfX$ is an uncertain variable $\calX$ which is partially random. This viewpoint leads to the introduction of o.p.m.s of the form
\eqns{
\oP_{\calX}(\varphi) = \sup_{\theta \in \Theta} f_{\uvtheta}(\theta) \int \varphi(x) p(x \param \theta) \d x = \sup\big\{ f_{\uvtheta}(\theta) p(\varphi \param \theta) \st \theta \in \Theta \big\}
}
for any function $\varphi$ on $\sfX$. Even though we define o.p.m.s via the values they give to functions, the credibility of the event $\calX \in B$ will be denoted by $\oP_{\calX}(B)$ instead of $\oP_{\calX}(\ind{B})$ for the sake of simplicity.

\begin{remark}
Defining probability distribution via functions or via sets can be proved to be equivalent \cite{Schwartz1973}. On the contrary, $\oP_{\calX}(\varphi)$ cannot be recovered from the knowledge of $\oP_{\calX}(B)$ for all $B \subseteq \sfX$ by lack of additivity so that defining the o.p.m.\ $\oP_{\calX}$ via the value it gives to functions on $\sfX$ is more general and this level of generality will be required.
\end{remark}

Throughout the article, it is shown that the loss of additivity incurred by considering o.p.m.s rather than probability distributions does not prevent from performing inference on $\uvtheta$ as well as on $\calX$. In some cases, the additional flexibility of possibility functions even proves useful to justify statistical techniques such as the empirical Bayes method \cite{Robbins1956, Carlin2010}. This aspect will be covered in Section~\ref{sec:parametrisedFamily} together with a more general treatment of o.p.m.s of the same form as $\oP_{\calX}$. This will be followed by the study of another type of o.p.m.\ that is closely related to fuzzy Dempster-Shafer theory \cite{Yen1990} in Section~\ref{sec:deterministicTransformation}.

As far as uncertain variables are concerned, a similar approach has been taken in \cite{DelMoral1999, DelMoral2000} in the context of optimal control where analogues of random variables are defined as \emph{control variables}. This formulation is related to idempotent analysis and $(\max,+)$-algebras \cite{Maslov1992, Butkovivc2010}. More recently, non-stochastic variables have been considered in \cite{Constantinou2017} with applications to causal inference. However, these non-stochastic variables are used to model known but varying parameters so that there is no associated notion of uncertainty.

The idea of using a different type of object for characterising the uncertainty about parameters dates back from Fisher \cite{Fisher1930} with the introduction of the notions of \emph{fiducial distribution} and \emph{fiducial inference}, extensions of which are still considered in the modern literature \cite{Hannig2009}. Fisher's motivation was to introduce a direct representation of the information in the parameter of a statistical model contained in the likelihood. Although the approach proposed in this article is fundamentally different from fiducial inference, our motivation for proposing an alternative representation of uncertainty is of the same nature. The differences between the proposed approach and fiducial inference will be illustrated in Section~\ref{sec:fiducial}. Some connections with Dempster-Shafer theory will also be studied in Sections~\ref{sec:Pearl} and \ref{sec:boxer} where some of the criticism of this theory by Pearl \cite{Pearl1990} and Gelman \cite{Gelman2006} are discussed.

\section{Uncertain variable}
\label{sec:uncertainVariable}

One of the fundamental concepts in the considered representation of uncertainty is the one of \emph{uncertain variable} which is used as an analogue of the concept of random variable in standard probability theory. We consider a sample space $\Omega_{\r}$ of probabilistic outcomes that is equipped with a probability distribution $\bbP$ in order to represent the involved randomness. We also consider another sample space $\Omega_{\u}$ which contains all the possible states of deterministic phenomena. There is no probability measure associated with $\Omega_{\u}$; instead, there is a reference element in $\Omega_{\u}$ that corresponds to the actual value of the considered parameters and which we denote $\omega_{\u}^{*}$. In general, the probability measure $\bbP$ might be conditional on the state $\omega_{\u}$ of the non-random phenomena described in $\Omega_{\u}$, in which case we write it as $\bbP(\bscdot \given \omega_{\u})$.

\begin{example}
\label{ex:unfairDie}
Consider an experiment where an operator is asked to
\begin{enumerate*}[label=\roman*)]
\item\label{it:dieExample_pick} pick any dice he fancies from a box containing $N$ unfair dice differently biased and
\item\label{it:dieExample_throw} throw it.
\end{enumerate*}
Part~\ref{it:dieExample_pick} of the experiment, i.e.\ the selection of the dice, can be modelled as deterministic and described by $\omega_{\u} \in \Omega_{\u} = \{1,\dots,N\}$ whereas part~\ref{it:dieExample_throw} can be modelled as random with the outcome in $\Omega_{\r} = \{1,\dots,6\}$ determined through the law $\bbP(\bscdot \given \omega_{\u})$ which depends on the selection $\omega_{\u}$ of the dice.
\end{example}

In Example~\ref{ex:unfairDie}, one could argue that the selection of the dice is random and/or that throwing the dice is deterministic. The viewpoint considered in this article, which is reminiscent of the frequentist approach, is that experiments that are expected to yield different outcomes with a certain frequency when repeated infinitely can be appropriately modelled as random whereas experiments that cannot easily be repeated or for which there is no regularity in the outcomes might be modelled as deterministic but uncertain. In the latter case, a possibility function is used to model the information available about the specific outcome of the phenomenon of interest. Parameters of statistical models can also be considered deterministic in general.

\subsection{Deterministic case}
\label{sec:deterministicUncertainVariable}

We start with the special case where only deterministic uncertainty is present and then introduce the general notion of uncertain variable in Section~\ref{sec:generalUncertainVariable}. A more detailed exposition of this case is given in \cite{Houssineau2019}.

\begin{definition}
A $\Theta$-valued deterministic uncertain variable is a mapping from $\Omega_{\u}$ to $\Theta$.
\end{definition}

Deterministic uncertain variables will be written as bold Greek letters, e.g.\ $\uvtheta$ or $\uvpsi$. We first note that the concept of realisation is not specific to randomness and applies straightforwardly in this context: a realisation $\theta$ of the deterministic uncertain variable $\uvtheta$ is simply the image $\uvtheta(\omega_{\u})$ for some $\omega_{\u} \in \Omega_{\u}$.

\begin{remark}
A deterministic uncertain variable does not have to be a measurable mapping; in fact, neither $\Theta$ nor $\Omega_{\u}$ are equipped with a $\sigma$-algebra.
\end{remark}

As opposed to random variables, whose laws are induced by the fundamental probability distribution on $\Omega_{\r}$, we will express our knowledge about a deterministic uncertain variable $\uvtheta : \Omega_{\u} \to \Theta$ directly via some o.p.m.\ on $\Theta$. In particular, we will consider o.p.m.s $\oP_{\uvtheta}$ of the form
\eqns{
\oP_{\uvtheta}(\varphi) = \sup_{\theta \in \Theta} \varphi(\theta) f_{\uvtheta}(\theta) = \sup \varphi \cdot f_{\uvtheta},
}
for any function $\varphi$ on $\Theta$, where $\varphi \cdot f_{\uvtheta}$ denotes the point-wise product between $\varphi$ and $f_{\uvtheta}$ and where $f_{\uvtheta}$ is a possibility function. The interpretation of the o.p.m.\ $\oP_{\uvtheta}$ is that $\oP_{\uvtheta}(B)$ is the credibility of the event $\uvtheta \in B$. For instance, we can say that $\uvtheta \in B$ happens almost surely (a.s.) if $\oP_{\uvtheta}(B^{\c}) = 0$. Possibility functions and the associated o.p.m.s are not induced by the corresponding deterministic uncertain variables but only represent what is known about them; hence, we say that $f_{\uvtheta}$ and $\oP_{\uvtheta}$ \emph{describe} $\uvtheta$.

The choice of the supremum/maximum as the main operator for outer measures representing deterministic uncertainty can be explained intuitively. Indeed, the use of a maximum shows that there is a particular point of interest in the domain of $f_{\uvtheta}$ which is the point where $f_{\uvtheta}$ is maximised. This is related to the fact that there is a single point of interest in $\Omega_{\u}$ which is the reference point $\omega_{\u}^{*}$.

It is demonstrated in \cite{Houssineau2019} that possibility functions and their properties allow for a law of large numbers and a central limit theorem to be derived for deterministic uncertain variables, leading to the definition of expected value and variance as
\eqns{
\bbE^*(\uvtheta) = \argmax_{\theta \in \Theta} f_{\uvtheta}(\theta) \AND \bbV^*(\uvtheta) = -\bigg( \dfrac{\d^2}{\d \theta^2} f_{\uvtheta}\big( \bbE^*(\uvtheta) \big) \bigg)^{-1},
}
where the expected value $\bbE^*(\uvtheta)$ is assumed to be a singleton when defining the variance $\bbV^*(\uvtheta)$, otherwise we set $\bbV^*(\uvtheta) = \infty$. These results confirm the relation between deterministic uncertainty and the supremum/maximum operators: the point of interest is the one that is the most likely to be corresponding to $\omega_{\u}^{*}$ and the uncertainty around this point is best described by a local quantity such as the proposed notion of variance.

To help the interpretation and definition of possibility functions, the associated o.p.m.\ can be seen as an upper bound for probability distributions. In that context, the credibility of an event $\uvtheta \in B$ can be regarded as the maximum subjective probability that one would be willing to attribute to that event when seeing $\uvtheta$ as a random variable. This indicates that the possibility function equal to $1$ everywhere is the least informative.

The analogues of standard operations for probability distributions can be deduced for possibility functions from the corresponding o.p.m.s: if $\uvtheta$ and $\uvpsi$ are two uncertain variables on $\Theta$ and $\Psi$ respectively which are jointly described by a possibility function $f_{\uvtheta,\uvpsi}$ on $\Theta \times \Psi$ then
\eqns{
f_{\uvpsi}(\psi) = \sup_{\theta \in \Theta} f_{\uvtheta,\uvpsi}(\theta, \psi), \qquad \psi \in \Psi,
}
is the marginal possibility function describing $\uvpsi$ and
\eqns{
f_{\uvtheta|\uvpsi}(\theta \given \psi) = \dfrac{ f_{\uvtheta,\uvpsi}(\theta,\psi) }{ f_{\uvpsi}(\psi) }, \qquad \theta \in \Theta,
}
is the conditional possibility function describing $\uvtheta$ given $\uvpsi$, which is defined \emph{for all credible $\psi \in \Psi$}, i.e.\ for all $\psi$ such that $f_{\uvpsi}(\psi) > 0$. For instance, the marginal possibility function $f_{\uvpsi}$ can be deduced from the o.p.m.\ $\oP_{\uvtheta}(\varphi) = \sup \varphi \cdot f_{\uvtheta,\uvpsi}$ by considering $\varphi = \phi \times \one_{\Theta}$ as follows
\eqns{
\oP_{\uvpsi}(\phi) = \oP_{\uvtheta}(\phi \times \one_{\Theta}) = \sup_{\psi \in \Psi} \phi(\psi) f_{\uvpsi}(\psi).
}
The expression of the conditional possibility function $f_{\uvtheta|\uvpsi}$ does not require the use probability density functions (p.d.f.s) even when $\Psi$ is uncountable since possibility functions can always be meaningfully evaluated point-wise. Analogues of standard concepts can also be easily introduced, for instance, $\uvtheta$ and $\uvpsi$ are said to be \emph{weakly independent} if there exist possibility functions $f_{\uvtheta}$ and $f_{\uvpsi}$ such that $f_{\uvtheta,\uvpsi}(\theta,\psi) = f_{\uvtheta}(\theta)f_{\uvpsi}(\psi)$ for any $(\theta,\psi) \in \Theta \times \Psi$. The possibility functions $f_{\uvtheta}$ and $f_{\uvpsi}$ describe $\uvtheta$ and $\uvpsi$ marginally. These types of operations on possibility functions are well known in possibility theory, see for instance \cite{Zadeh1978} and \cite{Dubois1981}. The independence property is interpreted as follows: $\uvtheta$ and $\uvpsi$ are weakly independent if the information we hold about $\uvtheta$ is unrelated to the one we hold about $\uvpsi$.

Many usual p.d.f.s can be easily transformed into possibility functions by simply renormalising them, for instance the function
\eqns{
\overline{\N}(\theta; \mu, \sigma^2) = \exp\bigg( -\dfrac{1}{2\sigma^2} (\theta-\mu)^2 \bigg), \qquad \theta \in \Theta \subseteq \bbR,
}
can be referred to as the normal/Gaussian possibility function with parameters $\mu \in \Theta$ and $\sigma > 0$. One important difference between the normal possibility function and its probabilistic analogue is that the former can be defined on a strict subset $\Theta$ of $\bbR$ without requiring normalisation. Renormalising p.d.f.s is suitable when these are conjugate priors in a Bayesian context since the induced possibility function will also be a conjugate prior.

It is generally simple to introduce new possibility functions, e.g.\ as exponentiated loss functions as studied in \cite{Bissiri2016}, since the condition that possibility functions have maximum one is easy to enforce. One can also deduce the possibility function describing a transformed uncertain variable via the following change of variable formula: if $\uvtheta$ is a deterministic uncertain variable on $\Theta$ described by $f_{\uvtheta}$ then, for any mapping $\zeta : \Theta \to \Psi$, the uncertain variable $\uvpsi = \zeta(\uvtheta)$ is also deterministic and can be described by the possibility function $f_{\uvpsi}$ characterised by
\begin{equation}
\label{eq:changeOfVariable}
f_{\uvpsi}(\psi) = \sup \big\{ f_{\uvtheta}(\theta) : \theta \in \zeta^{-1}[\psi] \big\}, \qquad \psi \in \Psi,
\end{equation}
where $\zeta^{-1}[A]$ is the inverse image of a subset $A$ of $\Psi$ by $\zeta$ and where we can ensure that the inverse image $\zeta^{-1}[\bscdot]$ is non-empty by assuming that $\zeta$ is surjective, otherwise the appropriate convention is $\sup \emptyset = 0$. There is no Jacobian term in \eqref{eq:changeOfVariable} since possibility functions are not densities.

The change of variable formula \eqref{eq:changeOfVariable} can be interpreted as pushing the information from $\Theta$ onto $\Psi$ via $\zeta$. The corresponding operation is therefore referred to as a \emph{push-forward}. If the information initially available is expressed as a possibility function $f_{\uvpsi}$ on $\Psi$, then it can also be useful to pull that information back onto $\Theta$ via $\zeta$, which differs in nature from the push-forward as soon as $\zeta$ is not bijective. The most fundamental instance of such a pull-back is to deduce the credibility of outcomes in $\Omega_{\u}$ based on the possibility function $f_{\uvpsi}$ describing a deterministic uncertain variable $\uvpsi$. Indeed, one can deduce from properties of o.p.m.s that the credibility of any $\omega_{\u} \in \Omega_{\u}$ is
$$
f(\omega_{\u}) = f_{\uvpsi}(\uvpsi(\omega_{\u})),
$$
which defines a possibility function $f$ on $\Omega_{\u}$ akin to the probability measure $\bbP$ on $\Omega_{\r}$, with the important difference that $f$ is induced by the available information on the parameter set $\Psi$ whereas $\bbP$ induces the law of random variables on $\Omega_{\r}$. The pull-back operation can also be used between two deterministic uncertain variables as follows: if $\uvpsi = \zeta(\uvtheta)$ for some given mapping $\zeta : \Theta \to \Psi$, and if the possibility function $f_{\uvpsi}$ is such that $\sup f_{\uvpsi} \circ \zeta = 1$ then one can deduce a possibility function for $\uvtheta$ as $f_{\uvtheta} = f_{\uvpsi} \circ \zeta$. The general case where $\sup f_{\uvpsi} \circ \zeta = 1$ does not necessarily hold is discussed in Appendix~\ref{sec:pullback}.

\begin{remark}
As opposed to possibility functions, probability distributions cannot be pulled back via non-bijective mappings without changes in the $\sigma$-algebra.
\end{remark}

In general, the possibility function $f$ on $\Omega_{\u}$ will be assumed to be induced by the joint credibility of all the deterministic uncertain variables of interest. In this way, the possibility function $f$ represents everything that is known about the considered deterministic uncertain variables. The expected value $\bbE^*(\uvtheta)$ of $\uvtheta$ can then be defined in a more canonical way as \cite{Houssineau2019}
\eqns{
\bbE^*(\uvtheta) = \uvtheta\Big( \argmax_{\omega_{\u} \in \Omega_{\u}} f(\omega_{\u}) \Big).
}
It is clear from this definition that this notion of expected value verifies $\bbE^*(\zeta(\uvtheta)) = \zeta(\bbE^*(\uvtheta))$ for any mapping $\zeta$, so that, noticing that the second derivative of $f_{\uvtheta}$ at the expected value is equal to the second derivative of $\log f_{\uvtheta}$ at the same point, the variance can be re-expressed as
\eqns{
\bbV^*(\uvtheta) = \bbE^*\bigg( - \dfrac{\d^2}{\d \theta^2} \log f_{\uvtheta}( \uvtheta) \bigg)^{-1},
}
that is, the variance is the inverse of a natural analogue of the Fisher information \cite{Houssineau2019}. This observation highlights the ability of possibility functions to represent information.

The applicability of possibility functions to complex sequential inference problems has been demonstrated in \cite{Houssineau2018, Houssineau2018_PHD, Ristic2019}. However, this approach does not allow for taking into account that some aspects of a statistical model might be sufficiently well-known to be characterised by probability distributions. The objective in the following section is to define the notion of uncertain variable which combines deterministic uncertain variables and random variables.

\subsection{General case}
\label{sec:generalUncertainVariable}

We have argued in the previous section that uncertain variables represent information about parameters instead of characterising random phenomena. The sample space $\Omega_{\u}$ contains a reference element $\omega^*_{\u}$ which yields the true value of the parameters of interest via the associated uncertain variables. Both the random aspect and the informational aspect coexist in standard probability theory, as illustrated in the following example.

\begin{example}
\label{ex:observedRandomVariable}
Consider a random variable $X$ with law $p_X$ on a space $\sfX$ and an event $X \in B$ for some subset $B$ of $\sfX$. The probability distribution $p_X$ characterises the randomness of $X$ and the event $X \in B$ relates to the information about a specific realisation of $X$. The latter relates to a reference element of $\Omega_{\r}$ of the same nature as $\omega^*_{\u}$. It should therefore be possible to define the conditional law $p_X(\bscdot \given X = \uvy)$ with $\uvy$ an uncertain variable on $\sfX$ described by a possibility function $f_{\uvy}$ which represents the available information about the realisation of $X$. We will show in this section that the law of $X$ and the information about $\uvy$ can be encapsulated in an o.p.m.\ $\bar{P}_{X,\uvy}$ characterised by
$$
\bar{P}_{X,\uvy}(\varphi) = \int \sup_{y \in \sfX} \big[ \varphi(x,y) f_{\uvy}(y) \big] p_X(\d x),
$$
for any function $\varphi$ on $\sfX \times \sfX$. Applying Bayes' rule to the o.p.m.\ $\bar{P}_{X,\uvy}$ yields
$$
\bar{P}_X(\phi \given X = \uvy) = \dfrac{\int \phi(x) f_{\uvy}(x) p_X(\d x)}{\int f_{\uvy}(x) p_X(\d x)},
$$
with $\phi(x) = \varphi(x,x)$ for any $x \in \sfX$. It holds that $\bar{P}_X(\bscdot \given X = \uvy)$ is equivalent to a probability measure which could be denoted $p_X(\bscdot \given X = \uvy)$. This is similar to Bayesian inference but with information about the realisation of $X$ being used in place of the likelihood; for instance, if $Y$ is a random variable with conditional law $p_Y(\bscdot \given x)$ yielding the observation $y$ then one can consider $f_{\uvy}(x) \propto p_Y(y \given x)$. If the possibility function describing $\uvy$ is of the form $f_{\uvy} = \ind{B}$ for some subset $B$ of $\sfX$, then $p_X(\bscdot \given X = \uvy) = p_X(\bscdot \given X \in B)$ as required. Alternatively, if loss functions are considered in the context of generalised Bayesian inference \cite{Bissiri2016} then $f_{\uvy}$ is the exponentiated negative loss function. 
\end{example}

Example~\ref{ex:observedRandomVariable} shows that uncertain variables can be used for representing information about realisations of random variables in addition to unknown parameters. The objective is now to formally interface the concept of deterministic uncertain variable with the one of random variable. The combination of these two concepts can easily be introduced as follows.

\begin{definition}
\label{def:uncertainVariable}
A $S$-valued uncertain variable is a mapping $\calX$ from $\Omega_{\u} \times \Omega_{\r}$ to $S$ such that $\calX(\omega_{\u}, \bscdot)$ is a random variable for any $\omega_{\u} \in \Omega_{\u}$.
\end{definition}

There are cases where $\calX$ and/or $\mathbb{P}$ will not depend on $\omega_{\u}$, yet both dependencies are needed in general. For instance, Example~\ref{ex:unfairDie} corresponds to the case where $\mathbb{P}$ does depend on $\omega_{\u}$ whenever the dice are not identically distributed; similarly, if the operator in this example has the choice between tossing a coin or a dice, then the underlying random variable has to depend directly on this choice since realisations will be in different sets, e.g.\ $\{\mathrm{H}, \mathrm{T}\}$ for the coin and $\{1,\dots,6\}$ for the dice.

Definition~\ref{def:uncertainVariable} is not directly useful since, in general, the law of $\calX(\omega_{\u}, \bscdot)$ would be parametrised by $\omega_{\u}$ and the possibility function describing $\calX(\bscdot, \omega_{\r})$ could be dependent on $\omega_{\r}$. The objective is however to have no explicit dependency on the sample space $\Omega_{\u} \times \Omega_{\r}$, as is standard for sample spaces, and we identify two special cases addressing this issue in the following sections.

\section{Special forms of uncertain variables}

\subsection{Parametrised family of random variables}
\label{sec:parametrisedFamily}

Considering a family of random variables parametrised by a fixed but unknown parameter is a very standard setting in Statistics, in particular in frequentist inference. The proposed approach suggests to additionally model the uncertainty about this unknown parameter via a possibility function. A form of generalised Bayesian inference follows from this modelling choice, as studied in \cite{Houssineau2019}.

In order to model this situation, we consider a $\Theta$-valued deterministic uncertain variable $\uvtheta$ and a mapping $Z : \Theta \times \Omega_{\r} \to \sfX$, which is simply a collection of random variables indexed by $\Theta$. These ingredients are then combined to form an uncertain variable $\calX$ of the form $Z(\uvtheta,\bscdot)$, i.e.\ such that $\calX(\omega_{\u},\omega_{\r}) = Z(\uvtheta(\omega_{\u}), \omega_{\r})$ for all $(\omega_{\u},\omega_{\r}) \in \Omega_{\u} \times \Omega_{\r}$. Although $Z(\theta, \bscdot)$ is a random variable for any given $\theta \in \Theta$, its law depends on $\omega_{\u}$ in general because of the potential dependency of $\bbP$ on $\omega_{\u}$. An additional formal condition needs to be introduced to ensure that the law of $Z(\theta, \bscdot)$ does not depend on $\omega_{\u}$. This condition relates to a form of sufficiency of the parameter $\theta$, which should guarantee that $\bbP(Z(\theta, \bscdot) \in B \given \theta, \omega_{\u})$ does not depend on $\omega_{\u}$ for any $B \subseteq \sfX$. The formal statement of this condition is relegated to Appendix~\ref{sec:sufficiency} and we henceforth assume that it is satisfied. We can therefore introduce the law $p_Z(\dotparam \theta)$ of $Z(\theta, \bscdot)$ for any $\theta \in \Theta$. The uncertain variable $\calX$ is then described by an o.p.m.\ $\bar{P}_{\calX}$ characterised by
$$
\bar{P}_{\calX}(\varphi) = \sup_{\theta \in \Theta} f_{\uvtheta}(\theta) \int \varphi(x) p_Z(x \param \theta) \d x,
$$
for any function $\varphi$ on $\sfX$. For instance, if $f_{\uvtheta}$ is the indicator of a subset $A$ of $\Theta$ and if $B$ is a subset of $\sfX$, then $\bar{P}_{\calX}(B)$ is the maximum probability assigned to the event $Z(\theta, \bscdot) \in B$ by the probability distribution $p_Z(\dotparam \theta)$ among all $\theta \in A$. If $\bar{P}_{\calX}$ is viewed as an upper bound for a subjective probability distribution $p_{\calX}$ on $\sfX$, then $p_{\calX}(B)$ can also be lower-bounded by
$$
p_{\calX}(B) = 1 - p_{\calX}(B^{\c}) \leq 1 - \bar{P}_{\calX}(B^{\c}),
$$
which is indeed the minimum probability that assigned to the event $Z(\theta, \bscdot) \in B$ for any $\theta \in A$. This type of upper and lower bounds are common with imprecise probability in general \cite{Walley1991}.

In order to learn about $\uvtheta$ via observed realisations of $\calX$ or to learn about $\calX$ via information about $\uvtheta$, the uncertain variables $\uvtheta$ and $\calX$ must be jointly described. The o.p.m.\ that is suitable for this purpose is
$$
\bar{P}_{\uvtheta,\calX}(\varphi) = \sup_{\theta \in \Theta} f_{\uvtheta}(\theta) \int \varphi(\theta, x) p_Z(x \param \theta) \d x,
$$
where the function $\varphi$ is now defined on $\Theta \times \sfX$. Bayes' rule can be applied to the o.p.m.\ $\bar{P}_{\uvtheta,\calX}$ in order to determine the conditional o.p.m.\ describing $\calX$ given $\uvtheta = \theta$:
\begin{equation}
\label{eq:condOpmThetaX}
\bar{P}_{\calX}(\varphi \given \theta) = \dfrac{\bar{P}_{\uvtheta,\calX}(\ind{\theta} \times \varphi)}{\bar{P}_{\uvtheta,\calX}(\ind{\theta} \times \ind{\sfX})} = \dfrac{f_{\uvtheta}(\theta) \int \varphi(x) p_Z(x \param \theta) \d x}{f_{\uvtheta}(\theta) \int p_Z(x \param \theta) \d x} = \int \varphi(x) p_Z(x \param \theta) \d x,
\end{equation}
for any function $\varphi$ on $\sfX$. Equation~\ref{eq:condOpmThetaX} shows that the conditional o.p.m.\ describing $\calX$ given $\uvtheta$ is in fact a probability distribution which we denote by $p_{\calX}(\bscdot \given \uvtheta)$. Since $p_{\calX}(\bscdot \given \uvtheta)$ is equal to $p_Z(\dotparam \uvtheta)$, we will henceforth use the former instead of the latter, which avoids references to $Z$.

Alternatively, assuming that $\sfX$ is countable, we compute the conditional o.p.m.\ describing $\uvtheta$ given $\calX = x$ and find
$$
\bar{P}_{\uvtheta}(\varphi \given x) = \sup_{\theta \in \Theta} \varphi(\theta) \dfrac{p_{\calX}(x \given \theta) f_{\uvtheta}(\theta)}{\sup_{\psi \in \Theta} p_{\calX}(x \given \psi) f_{\uvtheta}(\psi)} = \sup_{\theta \in \Theta} \varphi(\theta) f_{\uvtheta}(\theta \given x),
$$
for any function $\varphi$ on $\Theta$, where the second equation defines the posterior possibility function $f_{\uvtheta}(\bscdot \given x)$, which emphasises that $\bar{P}_{\uvtheta}(\bscdot \given x)$ is of a deterministic nature. The function $p_{\calX}(x \given \bscdot)$ is interpreted as a likelihood and the possibility function $f_{\uvtheta}$ is viewed as a prior. Inference in this context is studied in \cite{Houssineau2019} where a Bernstein-von Mises theorem is derived. 

We now aim to define a notion of expected value for the uncertain variable $\calX$. Since the conditional mean $\bbE(\calX \given \uvtheta)$ is a deterministic uncertain variable defined as
$$
\bbE(\calX \given \uvtheta) : \omega_{\u} \mapsto \int x\, p_{\calX}(x \given \uvtheta(\omega_{\u}))\d x,
$$
its expectation w.r.t.\ $\bbE^*(\bscdot)$ is meaningful. We therefore define the expected value of $\calX$ as
\begin{equation}
\label{eq:expectedValueThetaX}
\bbE^*(\calX) = \bbE^*\big( \bbE(\calX \given \uvtheta) \big).
\end{equation}
By the properties of $\bbE^*(\bscdot)$ when applied to deterministic uncertain variables, it holds that $\bbE^*(\calX) = \bbE(\calX \given \bbE^*(\uvtheta))$, that is, the expected value of $\calX$ is the mean of the random variable $Z(\theta, \bscdot)$ at the most credible parameter value $\theta = \bbE^*(\uvtheta)$. The definition of $\bbE^*(\calX)$ hints at a connection between uncertain variables and the empirical Bayes method \cite{Robbins1956, Carlin2010}. Indeed, the \emph{expected} law of $\calX$ can be defined as
$$
p^*_{\calX}(B) = \bbE^*\big( p_{\calX}(B \given \uvtheta) \big) = p_{\calX}(B \given \bbE^*(\uvtheta)), \qquad B \subseteq \sfX,
$$
where the parameter $\uvtheta$ is set to its most credible value as in the empirical Bayes method. To make this connection more explicit, we consider that $\calX$ is a hidden variable, and that we observe the realisation $y$ of another quantity $\calY$ dependent on $\calX$. Like $\calX$, the observed quantity $\calY$ is an uncertain variable even if it does not depend directly on $\uvtheta$ and we denote by $\bar{P}_{\calY}(\bscdot \given x)$ the conditional o.p.m.\ describing it for any $x \in \sfX$. We do not need to specify the nature of $\calY$ and simply denote by $L = \bar{P}_{\calY}(y \given \bscdot)$ the likelihood for the observation $y$. In this case, the o.p.m.\ jointly describing $\uvtheta$ and $\calX$ given $\calY = y$ is
\begin{align*}
\bar{P}_{\uvtheta,\calX}(\varphi \given y) & = \sup_{\theta \in \Theta} \dfrac{ f_{\uvtheta}(\theta) \int \varphi(\theta, x) L(x) p_{\calX}(x \given \theta)\d x}{\sup_{\psi \in \Theta} f_{\uvtheta}(\psi) \int L(x) p_{\calX}(x \given \psi)\d x} \\
& = \sup_{\theta \in \Theta} \dfrac{f_{\uvtheta}(\theta)\int L(x) p_{\calX}(x \given \theta)\d x}{\sup_{\psi \in \Theta} f_{\uvtheta}(\psi) \int L(x) p_{\calX}(x \given \psi)\d x} \dfrac{\int \varphi(\theta, x) L(x) p_{\calX}(x \given \theta)\d x}{\int L(x) p_{\calX}(x \given \theta) \d x} \\
& = \sup_{\theta \in \Theta} f_{\uvtheta}(\theta \given y) \int \varphi(\theta, x) p_{\calX}(x \given y, \theta) \d x
\end{align*}
for any function $\varphi$ on $\Theta \times \sfX$, where $f_{\uvtheta}(\bscdot \given y)$ is the posterior possibility function describing $\uvtheta$ given $\calY = y$ and where $p_{\calX}( \bscdot \given y, \theta)$ is the posterior distribution of $\calX$ given $(\calY,\uvtheta) = (y, \theta)$. The expected posterior distribution of $\calX$ is therefore characterised by
$$
p^*_{\calX}(B \given y) = \bbE^*\big( p_{\calX}( B \given y, \uvtheta) \Given \calY = y \big) = p_{\calX}\big( B \given y, \bbE^*(\uvtheta \given \calY = y) \big), \qquad B \subseteq \sfX.
$$
In particular, if $f_{\uvtheta}$ is uninformative, i.e.\ $f_{\uvtheta} = \ind{\Theta}$, then $\bbE^*(\uvtheta \given \calY = y)$ is the maximum likelihood estimate (MLE) of the parameter and $p^*_{\calX}(\bscdot \given y)$ is exactly the empirical Bayes posterior distribution. Empirical Bayes allows for greatly simplifying inference procedures by replacing a marginalisation over hyper-parameter with the MLE for the parameter. Applications where this approach has been successful include variable selection \cite{George2000} and microarray experiments \cite{Efron2001}. As opposed to empirical Bayes, the proposed approach also allows for taking into account the uncertainty about the unknown parameter $\uvtheta$ while still relying on optimisation rather than integration, e.g.\ the posterior credibility of an event $\calX \in B$ for some $B \subseteq \sfX$ is
$$
\bar{P}_{\calX}(B \given y) = \sup_{\theta \in \Theta} f_{\uvtheta}(\theta \given y) p_{\calX}(B \given y, \theta).
$$
A practical example of the use of this type of uncertain variable is given in Section~\ref{sec:empirical}.

\subsection{Deterministic transformation of a random variable}
\label{sec:deterministicTransformation}

The previous case was describing the situation where the distribution of a random variable depends on an unknown but fixed parameter. This is not the only form that uncertain variables can take. In a Bayesian context, and in particular in generalised Bayesian inference \cite{Bissiri2016}, it is common to update a prior distribution based on non-random data  \cite{Diaconis1982} as already discussed in Example~\ref{ex:observedRandomVariable}. More generally, we consider the situation where a random variable is first sampled and then transformed in a deterministic but uncertain manner. Such an experiment could represent the imprecise measurement of a physical quantity or the unknown choice of strategy of an opponent in a game of chance.

To model these experiments, we combine a $\sfX$-valued random variable $X$ with law $p_X$ and a mapping $\uvpsi : \Omega_{\u} \times \sfX \to \sfZ$ into an uncertain variable $\calZ$ of the form $\uvpsi(\bscdot, X)$, i.e.\ such that $\calX(\omega_{\u},\omega_{\r}) = \uvpsi(\omega_{\u}, X(\omega_{\r}))$ for all $(\omega_{\u},\omega_{\r}) \in \Omega_{\u} \times \Omega_{\r}$. For any given realisation $x$ of $X$, $\uvpsi(\bscdot, x)$ is an uncertain variable on $\sfZ$ and we denote by $f_{\uvpsi}(\dotparam x)$ the possibility function describing it. There are no conditions for introducing $f_{\uvpsi}(\dotparam x)$ in this case since it is defined directly on $\sfZ$. This case is referred to as a deterministic transformation of a random variable since it holds that $\calZ$ is defined by $\calZ = \psi(X)$ with $\psi = \uvpsi(\omega_{\u}^*, \bscdot)$ the true (but unknown) transformation from $\sfX$ to $\sfZ$. The o.p.m.\ $\oP_{X,\calZ}$ jointly describing $X$ and $\calZ$ is characterised by
$$
\oP_{X,\calZ}(\varphi) = \int \sup_{z \in \sfZ} \big[\varphi(x, z) f_{\uvpsi}(z \param x) \big] p_X(x) \d x,
$$
for any function $\varphi$ on $\sfX \times \sfZ$. We can proceed as before and determine the conditional o.p.m.\ describing $\calZ$ given $X = x$ as follows:
\begin{equation}
\label{eq:condOpmXZ}
\bar{P}_{\calZ}(\varphi \given x) = \dfrac{\sup_{z \in \sfZ} \big[\varphi(z) f_{\uvpsi}(z \param x) \big] p_X(x)}{\sup_{z \in \sfZ} \big[f_{\uvpsi}(z \param x)\big] p_X(x)} = \sup_{z \in \sfZ} \varphi(z) f_{\uvpsi}(z \param x),
\end{equation}
for any function $\varphi$ on $\sfZ$, so that $f_{\uvpsi}(\dotparam x)$ can be identified with the conditional possibility function describing $\calZ$ given $X = x$, which we denote by $f_{\calZ}(\bscdot \given x)$.

\begin{remark}
The supremum is measurable under general assumptions so that $\oP_{X,\calZ}$ is well defined, see \cite[Proposition 7.47]{Bertsekas1978}. We did not need to assume that $\sfX$ is countable in \eqref{eq:condOpmXZ} since $\bar{P}_{\calZ}(\bscdot \given x)$ can be defined as a Radon-Nikodym derivative.
\end{remark}

Although $\calZ$ is partially random, there might not be any available information regarding the dependency between $X$ and $\calZ$ so that $f_{\calZ}(\bscdot \given X)$ might not depend on $X$. In particular, if there is no information at all about the relation between $X$ and $\calZ$ then one has to set $f_{\calZ}(z \given X) = 1$ for any $z \in \sfZ$, which is indeed independent of $X$, as considered in Example~\ref{ex:observedRandomVariable}.

Uncertain variables of the same form as $\calZ$ are related to fuzzy Dempster-Shafer theory \cite{Yen1990} where $f_{\calZ}(\bscdot \given X)$ is the membership function of a fuzzy random set and where $p_X$ is related to the law of that fuzzy random set, referred to as a \emph{basic belief assignment} in that context. When viewing $f_{\calZ}(z \given x)$ as relating an observation $z \in \sfZ$ to a quantity of interest $x \in \sfX$, the relevant information is contained in the posterior o.p.m.\ describing $X$ given $\calZ = z$, characterised by
$$
\bar{P}_X(\varphi \given z) = \dfrac{\int \varphi(x) f_{\calZ}(z \given x) p_X(x) \d x}{\int f_{\calZ}(z \given x') p_X(x') \d x'} = \int \varphi(x) p_X(x \given z) \d x,
$$
for any function $\varphi$ on $\sfX$. The posterior distribution $p_X(\bscdot \given z)$ takes the same form as the usual Bayesian posterior, but with the likelihood defined as a possibility function. Although the posterior distribution does not depend on the way the likelihood is normalised, important quantities such as the marginal likelihood do depend on it and the asymptotic properties of estimators based on $p_X(\bscdot \given z)$ will also differ. This is relevant in the context of generalised Bayesian inference where the likelihood is based on a loss function which is non-random. The next example shows how more complex models can be constructed based on the two introduced types of uncertain variable.

\begin{example}
\label{ex:complexOpm}
Let $\uvtheta$ be a deterministic uncertain variable on $\Theta$ described by $f_{\uvtheta}$, let $\calX$ be an uncertain variable on $\sfX$ of the form $Z(\uvtheta, \bscdot)$ with conditional law $p_{\calX}(\bscdot \given \uvtheta)$ and let $\calZ$ be an uncertain variable on $\sfZ$ defined by
\begin{equation}
\label{eq:defZ}
\calZ(\omega_{\u}, \omega_{\r}) = \uvpsi\big(\omega_{\u}, Z(\uvtheta(\omega_{\u}), \omega_{\r})\big), \qquad (\omega_{\u}, \omega_{\r}) \in \Omega_{\u} \times \Omega_{\r},
\end{equation}
for some family $\uvpsi(\bscdot, x)$ of uncertain variables indexed by $x \in \sfX$. We denote by $f_{\calZ}(\bscdot \given X, \uvtheta)$ the conditional possibility function describing $\calZ$ given $X$ and $\uvtheta$. The interpretation of $\calZ$ is as follows: first an unknown parameter $\theta = \uvtheta(\omega_{\u}^*)$ is fixed, then a random variable $X(\theta,\bscdot)$ is realised to a value $x \in \sfX$ and, finally, $x$ is transformed in a deterministic but unknown fashion to $z = \psi(x)$ with $\psi = \uvpsi(\omega_{\u}^*, \bscdot)$. The o.p.m.\ $\oP_{\calV}$ describing the uncertain variable
$\calV = (\uvtheta, \calX, \calZ)$ in $\Theta \times \sfX \times \sfZ$ is characterised by
\eqns{
\oP_{\calV}(\varphi) = \sup_{\theta \in \Theta} f_{\uvtheta}(\theta) \int \sup_{z \in \sfZ} \big[ \varphi(\theta,x,z) f_{\calZ}(z \given x,\theta) \big]  p_{\calX}(x \given \theta) \d x
}
for any function $\varphi$ on $\Theta \times \sfX \times \sfZ$. The considered order between the supremum operators and the integral is the only possible one given the hierarchy between the different uncertain variables in \eqref{eq:defZ}. Marginal and conditional o.p.m.s can be introduced by evaluating the o.p.m.\ $\oP_{\calV}$ at a suitable function $\varphi$. For instance, the marginal o.p.m.\ describing $\calZ$ is characterised by
\eqns{
\oP_{\calZ}(\varphi) = \oP_{\calV}(\ind{\Theta} \times \ind{\sfX} \times \varphi)
}
for any function $\varphi$ on $\sfZ$ and the conditional o.p.m.\ describing $\calX$ given that $\calZ \in A$ for some $A \subseteq \sfZ$ is
\eqns{
\oP_{\calX}(\varphi \given \calZ \in A) = \dfrac{\oP_{\calV}(\ind{\Theta} \times \varphi \times \ind{A})}{ \oP_{\calZ}(A) }
}
for any function $\varphi$ on $\sfX$. It is possible to consider $A = \{z\}$ for some fixed $z \in \sfZ$ under weak conditions on $f_{\calZ}(\bscdot \given x, \theta)$ without having to assume that $\sfX$ is countable as in Section~\ref{sec:parametrisedFamily}. The intuition is that the realisation $x$ of $\calX$ cannot be determined to infinite precision if it lives, say, in $\bbR$; therefore, a small but systematic error will be made, which is modelled by $\uvpsi$, the measured quantity being $\calZ$. The model introduced in this example can therefore be useful to represent the measurement of an unknown physical quantity $\uvtheta$ through a random medium characterised by $\calX$ and by a finite-resolution sensor modelled by $\calZ$. This is for instance the case for radars which measure randomly-perturbed electro-magnetic signals via pixel-like resolution cells \cite{Skolnik1990}.
\end{example}

In general, o.p.m.s take an increasingly sophisticated form as the complexity of the considered statistical model grows. This aspect can however be managed by making suitable modelling decisions. For instance, if we consider a sequence $\calX_0,\dots,\calX_n$ of uncertain variables that alternate between the forms $\uvpsi(\bscdot, X)$ and $Z(\uvtheta, \bscdot)$, then the complexity of the corresponding o.p.m.s will grow with $n$. However, if $\calX_0$ is equivalent to a deterministic uncertain variable $\uvtheta$ describing the unknown but fixed initial value of a quantity of interest and if $\calX_1,\dots,\calX_n$ are uncertain variables of the form $Z_n(\uvtheta, \bscdot)$ modelling the subsequent values of that (randomly evolving) quantity, then the form of the corresponding o.p.m.s will remain simple.

\section{O.p.m.s as a practical representation of uncertainty}
\label{sec:practical}

In this section, we discuss a number of examples which, in spite of their simplicity, illustrate fundamental aspects of the proposed representation of uncertainty.

\subsection{Empirical Bayes and the James-Stein estimator}
\label{sec:empirical}

The standard setting for the James-Stein estimator \cite{Stein1956, James1961} is the following: a vector $y \in \bbR^d$ of observations is received, which is assumed to be the realisation of a random variable $Y$ with law $\N(X, \sigma^2 I_d)$, where $I_d$ the identity matrix of dimension $d$ and where $X$ another normal random variable with mean $\mu \in \bbR^d$ and unknown variance. We want to model this unknown variance with a deterministic uncertain variable $\uvs$ in $(0, \infty)$ described a priori by the uninformative possibility function $f_{\uvs}$, i.e.\ $f_{\uvs}(s) = 1$ for any $s > 0$. In this context, both $X$ and $Y$ are uncertain variables such that the conditional law of $X$ given $\uvs$ is $p_X(\bscdot \given \uvs) = \N(\mu, \uvs I_d)$ and the conditional law of $Y$ given $X$ is $\N(X, \sigma^2 I_d)$. The proposed model is very close to the one usually adopted for the James-Stein estimator \cite{Lehmann2006}; yet, the fact that the variance $\uvs$ is modelled as an uncertain variable allows for modelling the complete absence of prior information about this parameter. This leaves the opportunity to completely ignore the prior information contained in the mean $\mu$ of $X$ if this information does not fit the data since there is no objection against $\uvs$ being arbitrary large a priori. Standard calculations yield that the posterior law of $X$ given $\uvs$ is normal with mean $\hat{\mu}(\uvs) = (\sigma^2 \mu + \uvs y)/(\sigma^2 + \uvs)$ and variance $\sigma^2 \uvs/(\sigma^2 + \uvs)I_d$ and that the posterior possibility function describing $\uvs$ is
$$
f_{\uvs}( s \given y ) = \bigg( \dfrac{\| y- \mu \|_2^2}{d(\sigma^2 + s)} \bigg)^{d/2} \exp\bigg( \dfrac{d}{2} - \dfrac{\| y- \mu \|_2^2}{2(\sigma^2 + s)} \bigg).
$$
The function $f_{\uvs}( \bscdot \given y )$ can be interpreted as a \emph{shifted inverse-gamma possibility function}
$$
\overline{\mathrm{SIG}}(s; \alpha, \beta, \sigma^2) = \bigg( \dfrac{\beta}{\alpha(\sigma^2 + s)} \bigg)^{\alpha} \exp\bigg( \alpha - \dfrac{\beta}{(\sigma^2 + s)} \bigg),
$$
with $\alpha = d/2$ and $\beta = \| y- \mu \|_2^2/d$.

In general, if a deterministic uncertain variable $\uvtheta$ is described by a shifted inverse-gamma possibility function $\overline{\mathrm{SIG}}(\alpha, \beta, \sigma^2)$, then its expected value and variance are defined as
$$
\bbE^*(\uvtheta) = \dfrac{\beta}{\alpha} - \sigma^2 \AND \bbV^*(\uvtheta) = \dfrac{\alpha^3}{\beta^2}
$$
if $\beta/\alpha > \sigma^2$ and $\bbE^*(\uvtheta) = \bbV^*(\uvtheta) = 0$ otherwise. The shifted inverse-gamma possibility function is a conjugate prior for normal likelihoods with a partially unknown variance and the inverse-gamma possibility function can be recovered when $\sigma^2 = 0$.

\begin{remark}
The existence of a conjugate prior for a likelihood with partially unknown variance helps to compensate for the fact that a small deterministic measurement error must be added when observing the realisation of an uncertain variable of the form $Z(\uvtheta, \bscdot)$ on an uncountable space, as discussed in Example~\ref{ex:complexOpm}. Indeed, the presence of two likelihood functions, one defined as a normal possibility function with known variance and one defined as a normal probability distribution with unknown location and variance, does not prevent from deriving analytical expressions of the posterior possibility function for the unknown location via the shifted inverse-gamma possibility function.
\end{remark}

Coming back to the situation of interest, it holds that the posterior expected value $\bbE^*(\uvs \given y)$ is equal to the MLE since there is no prior information about $\uvs$. The expected value of the uncertain variable $X$ can then be expressed as
$$
\bbE^*(X \given y) = \bbE^*\big( \bbE(X \given y, \uvs) \Given y \big)= \hat{\mu}\big( \bbE^*(\uvs \given y) \big) = \max\big\{0, 1 - \sigma^2\alpha/\beta\big\}(y - \mu) + \mu,
$$
which is a modified version of the James-Stein estimator where the MLE for the variance is used instead of an unbiased estimator \cite{Lehmann2006}.

Alternatively, one could consider a model where $\uvtheta$ is a deterministic uncertain variable such that the conditional law of $Y$ given $\uvtheta$ is $\N(\uvtheta, \sigma^2 I_d)$, with $\uvtheta$ described by the normal possibility function $f_{\uvtheta}(\theta) = \overline{\N}(\theta;\mu,\tau^{-1})$, $\theta \in \bbR^d$, for some given precision $\tau \geq 0$. It is not meaningful in this case to learn the precision $\tau$ since $\overline{\N}(\mu,\tau^{-1})$ represents what is known about $\uvtheta$ and one cannot infer the uncertainty in the information one holds. The posterior expected value of $\uvtheta$ is $\bbE^*(\uvtheta \given y) = \hat{\mu}(\tau^{-1})$ which is the same estimator as with a normal probabilistic model with known precision $\tau$, however the case $\tau = 0$ yields an improper prior in the probabilistic case.

\subsection{Fiducial inference}
\label{sec:fiducial}

Fiducial inference \cite{Fisher1930,Fisher1935} aims to make probabilistic statements about the parameter of a statistical model based on the received observation without relying on a prior probability distribution. Although this motivation is very similar to the one of the proposed approach, the results differ as illustrated in this section via a simple example.

\begin{example}[Fiducial inference with a uniform likelihood]
\label{ex:fiducial}
We receive $n$ observations $y_1,\dots,y_n$ which are assumed to be realisations of the i.i.d.\ random variables $Y_1,\dots,Y_n$ with uniform distribution on the interval $[0,\theta^*]$, for some unknown $\theta^* > 0$. A sufficient statistics for $\theta^*$ is $S = \max\{Y_1, \dots, Y_n\}$ with cumulative distribution function $F_S(s) = (s/\theta^*)^n$. Given a possible value $\theta > 0$ of $\theta^*$, if we fix $\alpha \in (0,1)$, the event $S > \theta(1-\alpha)^{1/n}$ has probability $\alpha$. Fiducial inference suggests that this statement can be reversed to ``the event $\theta < s (1-\alpha)^{-1/n}$ has probability $\alpha$'' where $s = \max\{y_1, \dots, y_n\}$ is the realisation of $S$ and where $\theta$ is now seen as a random variable.
\end{example}

\begin{example}
We consider the same setting as Example~\ref{ex:fiducial} but with the uncertainty about $\theta^*$ modelled by a deterministic uncertain variable $\uvtheta$ on $(0,\infty)$ with uninformative possibility function $f_{\uvtheta}$, i.e.\ $f_{\uvtheta}(\theta) = 1$ for all $\theta \in \Theta$. The likelihood is $L(\theta) = \theta^{-n}\prod_{i=1}^n \ind{[0,\theta]}(y_i)$ so that the posterior possibility function describing $\uvtheta$ given $y_1,\dots,y_n$ is
$$
f_{\uvtheta}(\theta \given y_1,\dots,y_n) = \dfrac{L(\theta) f_{\uvtheta}(\theta)}{\sup_{\psi > 0} L(\psi) f_{\uvtheta}(\psi)} = \dfrac{L(\theta)}{\sup_{\psi > 0} L(\psi)} = \dfrac{s^n}{\theta^n} \ind{[s,\infty)}(\theta),
$$
for any $\theta > 0$, with $s = \max\{y_1, \dots, y_n\}$ as before. The posterior possibility function $f_{\uvtheta}(\cdot \given y_1,\dots,y_n)$ is the renormalised version of the Pareto distribution, which can be made uninformative when $s = n = 0$ as opposed to its probabilistic counterpart. The posterior credibility of any event of the form $\uvtheta < s + \epsilon$ with $\epsilon > 0$ is
$$
\bar{\bbP}( \uvtheta < s + \epsilon \given y_1,\dots,y_n ) = \sup \Big\{ \dfrac{s^n}{\theta^n} \ind{[s,\infty)}(\theta) : \theta < s + \epsilon \Big\} = 1,
$$
where $\bar{\bbP}( \bscdot \given y_1,\dots,y_n)$ is the o.p.m.\ induced by the posterior possibility function $f_{\uvtheta}(\bscdot \given y_1,\dots,y_n)$ on $\Omega_{\u}$. Indeed, there is no objection against $\uvtheta$ being arbitrarily close to $s$ as long as $\uvtheta \geq s$. This does not mean that there is no additional information about $\uvtheta$, we know for instance that the posterior credibility for $\uvtheta$ to be greater than $2s$ is equal to $2^{-n}$, which vanishes when $n$ tends to infinity as required.
\end{example}

This example illustrates how the proposed approach enables the ``inversion'' of the likelihood without requiring prior information as envisioned by Fisher. As opposed to standard fiducial inference, the statistical framework based on o.p.m.s does not require the existence or uniqueness of sufficient statistics \cite{Dempster1963} and simply follows and extends the principles of Bayesian inference.

Another important aspect of this example is that an uninformative probabilistic prior taken to be constant on $(0,\infty)$ would yield an improper posterior in the standard Bayesian inference framework when a single observation is received, i.e.\ when $n = 1$. Defining the least informative proper prior distribution is often possible but requires more advanced calculations, see for instance \cite{Berger2009}. Generalised fiducial inference \cite{Hannig2009} provides solutions where the posterior quantification of uncertainty is also a probability distribution.

\subsection{The criticism of Dempster-Shafer theory by Pearl \cite{Pearl1990}}
\label{sec:Pearl}

Through several examples, it is shown in \cite{Pearl1990} how belief functions might not always be appropriate to represent
\begin{enumerate*}[label=\roman*)]
\item incomplete knowledge,
\item belief updating and
\item evidence pooling.
\end{enumerate*}
Belief functions are defined as follows: let $m : 2^{\sfX} \to [0,1]$ be a basic belief assignment on a finite set $\sfX$ with $2^{\sfX}$ the power set of $\sfX$, i.e.\ $m$ is a probability mass function on the subsets of $\sfX$ such that $m(\emptyset) = 0$, then the corresponding belief function $\Bel : 2^{\sfX} \to [0,1]$ and plausibility function $\Pl : 2^{\sfX} \to [0,1]$ are defined by
$$
\Bel(A) = \sum_{B \subseteq A} m(B) \AND \Pl(A) = \sum_{B: B \cap A \neq \emptyset} m(B)
$$
for any $A \subseteq \sfX$. Plausibility functions can be identified with the o.p.m.s describing uncertain variables $\calZ$ of the form $\uvpsi(\bscdot, X)$ where $X$ is a random variable in $2^{\sfX}$ with law $m$ and where the deterministic uncertain variable $\uvpsi(\bscdot, x)$ on $\sfZ = \sfX$ is described by $f_{\uvpsi}(z ; A) = \ind{A}(z)$ for any $A \subseteq \sfX$. Indeed, in this case, we have
$$
\oP_{\calZ}(\varphi) = \sum_{B \subseteq \sfX} \sup_{z \in B} \varphi(z) m(B) 
$$
for any function $\varphi$ on $\sfX$, which yields $\oP_{\calZ}(A) = \Pl(A)$ for any $A \subseteq \sfX$. Dempster's conditioning is then Bayes' theorem for o.p.m.s and Dempster's rule of combination is equivalent to conditioning an o.p.m.\ describing two uncertain variables $\calZ$ and $\calZ'$, of the form assumed here, on the event $\calZ = \calZ'$. This connection highlights the differences between the two approaches:
\begin{enumerate*}[label=\roman*)]
\item via the introduction of uncertain variables, defining models and assumptions in the context o.p.m.s is close to the standard statistical approach and is therefore facilitated and
\item o.p.m.s can take several easily-interpretable forms which are adapted to different situations.
\end{enumerate*}
We illustrate this latter point by reconsidering Example 1 of \cite{Pearl1990} in the context of o.p.m.s as follows. 

\begin{example}
\label{ex:threeExlusiveEvents}
Let $E_1$, $E_2$ and $E_3$ be three exclusive events (in $\sfX$) for which the only available information is
$$
0 \leq \bbP(E_i) \leq 1/2, \qquad i = 1,2,3.
$$
It can be shown that no belief function induces the bounds $\Bel(E_i) = 0$ and $\Pl(E_i) = 1/2$ for $i = 1,2,3$. We consider that $\sfX$ is the set defined by the union of the disjoint sets $E_1$, $E_2$, $E_3$ and $E_4 \doteq E_1^{\c} \cap E_2^{\c} \cap E_3^{\c}$, which form a partition of $\sfX$ by construction. In order to model this situation, we introduce a deterministic uncertain variable $\uvtheta$ on a set $\Theta$ which is assumed to be the standard 3-simplex, i.e.\
$$
\Theta = \{ (\theta_1,\dots,\theta_4) \in [0,1]^4 : \theta_1 + \dots + \theta_4 = 1 \}.
$$
We also introduce an uncertain variable $\calX$ on $\sfX$ of the form $Z(\uvtheta, \bscdot)$, such that $p_Z(E_i \param \uvtheta) = \uvtheta_i$ for any $i \in \{1,2,3\}$. The information given above can then be encoded in the possibility function $f_{\uvtheta}$ defined as the indicator of the subset $\{ \theta \in \Theta : \theta_i \leq 1/2 \et i \in \{1,2,3\} \}$ of $\Theta$. The marginal o.p.m.\ describing $\calX$ is such that
$$
\oP_{\calX}(B) = \sup_{\theta \in \Theta} f_{\uvtheta}(\theta) p_Z(B \param \theta),
$$
for any subset $B$ of $\sfX$. It follows that
\begin{align*}
\oP_X(E_i) & = \sup_{\theta \in \Theta} f_{\uvtheta}(\theta) \theta_i = 1/2, \qquad i = 1,2,3 \\
1 - \oP_X(E_i^{\c}) & = 1 - \sup_{\theta \in \Theta} f_{\uvtheta}(\theta) (\theta_j + \theta_k + \theta_l) = 0, \qquad \{i,j,k,l\} = \{1,2,3,4\},
\end{align*}
as required.
\end{example}

In Example~\ref{ex:threeExlusiveEvents}, it is the ability to consider uncertain variables $\calX$ of the form $Z(\uvtheta, \bscdot)$ that brings additional capabilities for modelling imprecise probabilities. This is consistent with the observation made above that it is the other main type of uncertain variables, i.e.\ the ones for which randomness comes first, that are related to (fuzzy) Dempster-Shafer theory. Cases where only marginal or conditional probabilities of events are known are then considered in \cite{Pearl1990}, which can be solved in a similar way as Example~\ref{ex:threeExlusiveEvents}. These results are not surprising since $\oP_{\calX}$ is an upper bound for (possibly objective) probabilities when $f_{\uvtheta}$ is an indicator function; yet, they show how o.p.m.s can represent different types of information. In \cite{Pearl1990}, Pearl concludes that belief functions can be useful as tools for eliciting evidence from natural language statements in order to update prior probability distributions. This advantage is preserved in the proposed approach via the type of model considered in Example~\ref{ex:observedRandomVariable}, see for instance \cite{Bishop2018}. In addition, more complex o.p.m.s such as the one considered in Example~\ref{ex:complexOpm} allow for modelling uncertain prior distributions.

\subsection{Ordering and causality}

We consider a deterministic uncertain variable $\uvtheta$ on $\Theta$ described by a possibility function $f_{\uvtheta}$ and a random variable $X$ on $\sfX$ with law $p_X$. These quantities are unrelated since $f_{\uvtheta}$ does not depend on $X$ and $p_X$ is not conditioned on $\uvtheta$. In spite of the fact that this is a very simple situation, an ambiguity arises when considering uncertain variables of the form $T(\uvtheta, X)$ with $T$ some given mapping on $\Theta\times\sfX$. Indeed, there is no natural order in this case and yet the supremums and integrals do not commute in general. We first emphasise that, as long as $\varphi$ is a non-negative function, it holds that $\sup_{\theta \in \Theta} \varphi(\theta, x) f_{\uvtheta}(\theta) = \| \varphi(\bscdot, x) \cdot f_{\uvtheta} \|_{\infty}$ for any $x \in \sfX$, with $\|\bscdot\|_{\infty}$ the uniform norm, which implies that
\eqns{
\bigg\| f_{\uvtheta} \cdot \int \varphi(\bscdot,x) p_X(x) \d x \bigg\|_{\infty} \leq  \int \| \varphi(\bscdot, x) \cdot f_{\uvtheta} \|_{\infty} p_X(x) \d x
}
so that the right hand side is interpreted as being less informative than the left hand side, the latter being therefore preferred in general. There is however a more fundamental aspect to this question of ordering, as illustrated in the following example.

\begin{example}
\label{ex:sumRandomDeterministic}
Assume that $\Theta = \sfX = \{1,\dots,6\}$, that is $X$ corresponds to a dice-rolling experiment and $\uvtheta$ is simply an unknown fixed integer between $1$ and $6$ (e.g.\ asking a player to pick a number between $1$ and $6$ as part of a game). Assume additionally that $\calX = \uvtheta + X$, i.e.\ we simply add the two obtained integers, that the dice is fair and that there is no information about $\uvtheta$, i.e.\ $f_{\uvtheta}(\theta) = 1$ for any $\theta \in \Theta$. There are two possible o.p.m.s corresponding to two different orderings between the supremum and the integral:
\eqns{
\oP_{\uvtheta,X}(\varphi) = \max_{\theta \in \{1,\dots,6\}} \dfrac{1}{6} \sum_{n = 1}^6 \varphi(\theta + n) \AND \oP_{X,\uvtheta}(\varphi) = \dfrac{1}{6} \sum_{n = 1}^6 \max_{\theta \in \{1,\dots,6\}} \varphi(\theta + n).
}
The credibility of any given sum $\calX$ between $2$ and $12$ is reported in Table~\ref{tbl:credibilitySums}. These credibilities significantly differ and indeed correspond to different experiments: $\oP_{\uvtheta,X}$ models the case where the value of $\uvtheta$ is fixed in advance so that, say, $\calX = 7$ is not more credible than $\calX = 2$ whereas $\oP_{X,\uvtheta}$ models the case where the value of $\uvtheta$ can be chosen after the rolling of the dice, with realisation $x \in \sfX$, in which case the value $7$ can be obtained systematically by choosing $\uvtheta = 7 - x$. If $\uvtheta$ models the choice of a player then the order corresponding to $\oP_{X,\uvtheta}$ models the case where that player might have observed the outcome of the dice roll before picking her own number. The credibilities associated with $\oP_{\uvtheta,X}$ for different values of $\calX$ neither sum to $1$ nor have maximum $1$; this is because the corresponding uncertain variable is neither deterministic nor random, which is fine as long as $\oP_{\uvtheta,X}(\{2,\dots,12\}) = 1$.
\end{example}

\begin{table}
\centering
\caption{Credibilities of the different values of $\calX$ under the two o.p.m.s defined in Example~\ref{ex:sumRandomDeterministic}.}
\label{tbl:credibilitySums}
\renewcommand{\arraystretch}{1.5}
\begin{tabular}{l|ccccccccc}
sum ($s$) & 2 & 3 & \dots & 6 & 7 & 8 & \dots & 11 & 12 \\
\hline
$\oP_{\uvtheta,X}(s)$ & 1/6 & 1/6 & \dots & 1/6 & 1/6 & 1/6 & \dots & 1/6 & 1/6 \\
$\oP_{X,\uvtheta}(s)$ & 1/6 & 2/6 & \dots & 5/6 & 1   & 5/6 & \dots & 2/6 & 1/6 \\
\end{tabular}
\end{table}

One aspect of Example~\ref{ex:sumRandomDeterministic} is still unclear: even if the integer $\uvtheta$ was chosen after the dice roll, there should be a notion of independence that lets us model that $\uvtheta$ was chosen independently of $X$; this should imply that the order between $\uvtheta$ and $X$ is irrelevant. Such a notion of independence is not needed if $\uvtheta$ comes first since an uncertain variable of the form $Z(\uvtheta, \bscdot)$ is \emph{characterised} by $p_Z(\dotparam \uvtheta)$ so that setting $p_Z(\dotparam \uvtheta) = p_X$ is sufficient to model that the random variable $Z(\theta, \bscdot)$ is in fact independent of $\theta$. In the opposite situation where $X$ comes first, making $f_{\uvpsi}(\dotparam x)$ independent of $x$ is not sufficient to ensure the dependence of the deterministic uncertain variable $\uvpsi(\bscdot, x)$ on $x$ in general. The independence is only guaranteed in the degenerate case where there is no more uncertainty about $\uvpsi$, that is $f_{\uvpsi}(z ; x) = \ind{z^*}(z)$ for some given $z^* \in \sfZ$. There is however a stronger notion of independence which makes sense in this context. For the sake of simplicity, we discuss this notion for two deterministic uncertain variables.

Recalling that the notion of weak independence introduced above only relates to the available knowledge about deterministic uncertain variables, we aim to introduce a stronger notion of independence that is directly related to the variables themselves. In particular, if $f_{\uvtheta,\uvpsi}$ jointly describes the deterministic uncertain variables $\uvtheta$ and $\uvpsi$, then the fact that $f_{\uvtheta,\uvpsi}(\theta,\psi) = f_{\uvtheta}(\theta)f_{\uvpsi}(\psi)$ only means that the information we have about $\uvtheta$ is not related to the one we have about~$\uvpsi$. Overall, $f_{\uvtheta,\uvpsi}$ is only weakly related to $\uvtheta$ and $\uvpsi$ as mappings on $\Omega_{\u}$.

Intuitively, $\uvtheta$ can be seen as (strongly) independent of $\uvpsi$ if fixing the value of $\uvpsi$ does not affect the behaviour of $\uvtheta$. To state this more formally, we introduce the deterministic uncertain variable $\uvtheta\given\uvpsi = \psi$ as the mapping
\eqnsa{
(\uvtheta\given\uvpsi = \psi) : \uvpsi^{-1}[\psi] & \to \Theta \\
\omega_{\u} & \mapsto \uvtheta(\omega_{\u})
}
that is, $\uvtheta\given\uvpsi = \psi$ is the restriction of $\uvtheta$ to $\uvpsi^{-1}[\psi] \subseteq \Omega_{\u}$. We can then say that $\uvtheta$ and $\uvpsi$ are \emph{strongly independent} if $\uvtheta\given\uvpsi = \psi$ and $\uvtheta$ have the same co-domain for any $\psi \in \Psi$. This definition follows the same motivation as the notion of conditional independence introduced in \cite{Constantinou2017} for non-stochastic variables. For example, if $\Omega_{\u} = \Theta \times \Psi$, $\uvtheta(\theta, \psi) = \theta$ and $\uvpsi(\theta,\psi) = \psi$ then $\uvtheta$ and $\uvpsi$ are strongly independent as expected.

The two notions of independence are partially related: if we know about the dependence of $\uvtheta$ on $\uvpsi$ then we can take it into account in the corresponding possibility function $f_{\uvtheta,\uvpsi}$ as illustrated in the example below. However, no new information is brought by strong independence since the fact that $\uvtheta$ is supported by $\Theta$ for any realisation $\psi$ of $\uvpsi$ is uninformative in terms of possibility function. Indeed, $f_{\uvtheta,\uvpsi}$ only relates to the true values $\uvtheta(\omega^*_{\u})$ and $\uvpsi(\omega^*_{\u})$ and not to the way $\uvtheta$ and $\uvpsi$ behave as functions on $\Omega_{\u}$. Conversely, there is no way to enforce the independence between $\uvtheta$ and $\uvpsi$ via possibility functions, as illustrated in Example~\ref{ex:sumRandomDeterministic}.

\begin{example}
If we are given the information that there exists a mapping $\zeta : \Theta \to \Psi$ such that $\uvpsi = \zeta(\uvtheta)$ then we can compute the conditional possibility function
$$
f_{\uvtheta,\uvpsi}(\theta,\psi \given \uvpsi = \zeta(\uvtheta)) = \ind{\zeta(\theta)}(\psi) \dfrac{f_{\uvtheta,\uvpsi}(\theta,\psi)}{\sup \{ f_{\uvtheta, \uvpsi}(\theta', \psi) : \theta' \in \Theta, \zeta(\theta')= \psi \}},
$$
for any $(\theta,\psi) \in \Theta \times \Psi$. In particular, if $\zeta$ is bijective then $f_{\uvtheta,\uvpsi}(\bscdot \given \uvpsi = \zeta(\uvtheta))$ is simply the indicator of the graph of $\zeta$, defined as $\{(\theta,\psi) \in \Theta \times \Psi : \psi = \zeta(\theta)\}$. The same operation in the context of probability theory would not be well defined for the reasons behind the Borel-Kolmogorov paradox \cite{Jaynes2003}. Such an operation can however be relevant in practice such as in the context of Bayesian melding \cite{Poole2000}.
\end{example}

\subsection{The boxer, the wrestler, and the coin flip}
\label{sec:boxer}

In order to further illustrate the concept of uncertain variable, we consider the interesting example introduced in \cite{Gelman2006} and show how the framework introduced in this article can explain some seemingly counter-intuitive results.

We assume that $\Theta = \sfX = \{0,1\}$, that the deterministic uncertain variable $\uvtheta$ models the ``hypothetical fight to the death between the world's greatest boxer and the world's greatest wrestler'' and that the random variable $X$ models tossing a fair coin. More specifically, the value $1$ is associated with the outcomes ``heads'' and ``the boxer wins''. Using the formalism of Dempster-Shafer theory, Gelman conditions on the event that the coin flip and the fight have the same outcome (in $\{0,1\}$) and noting that the posterior uncertainty about the fight is a probability distribution, concludes that the fight has counter-intuitively become random.

There are two possible orderings because of the absence of knowledge about which of $\uvtheta$ or $X$ came first, and we start with the case where $X$ comes first. It is useful to consider a slightly generalised setting where the law $p_X$ of $X$ is arbitrary and where $\uvtheta$ is described by any given possibility function $f_{\uvtheta}$. The o.p.m.\ jointly describing $X$ and $\uvtheta$ is
\begin{align*}
\oP_{X,\uvtheta}(\varphi) & = p_X(0) \max \big\{ \varphi(0,0)f_{\uvtheta}(0), \varphi(0,1)f_{\uvtheta}(1) \} \\
& + p_X(1) \max\big\{ \varphi(1,0)f_{\uvtheta}(0), \varphi(1,1)f_{\uvtheta}(1) \big\}.
\end{align*}
for any function $\varphi$ on $\{0,1\}^2$. We then introduce the conditional o.p.m.\ $\oP_{\uvtheta}(\bscdot \given x)$ defined by
$$
\oP_{\uvtheta}(\varphi(x, \bscdot) \given x) = \max \big\{ \varphi(x, 0) f_{\uvtheta}(0) \et \varphi(x, 1) f_{\uvtheta}(1) \big\},
$$
for any function $\varphi$ on $\{0,1\}^2$, where we let $\varphi$ depend on $x$ in order to maintain generality. The o.p.m.\ $\oP_{X,\uvtheta}$ can then be expressed as $\oP_{X,\uvtheta}(\varphi) = p_X( \oP_{\uvtheta}(\varphi \given \bscdot) )$. It follows that, when conditioned on an event $E$ in $\sfX \times \Theta$, the marginal o.p.m.\ $\oP_X$ becomes
\begin{equation}
\label{eq:XGivenE}
p_X(\phi \given E) = \dfrac{p_X( \phi \cdot \oP_{\uvtheta}(E \given \bscdot) )}{p_X( \oP_{\uvtheta}(E \given \bscdot) )},
\end{equation}
for any function $\phi$ on $\{0,1\}$. The term $p_X(\bscdot \given E)$ is the posterior distribution describing $X$ given $E$. It is necessarily a probability distribution in this case. Equation~\ref{eq:XGivenE} shows that $\oP_{\uvtheta}(\bscdot \given x)$ naturally acts as a likelihood for $X$. When considering Gelman's example, i.e.\ $f_{\uvtheta}$ is uninformative, the coin is fair and $E = \{X = \uvtheta\}$, we obtain
$$
p_X(\phi \given X = \uvtheta) = \dfrac{1}{2} \big( \phi(0) + \phi(1) \big) = p_X(\phi),
$$
for any function $\phi$ on $\{0,1\}$. This posterior distribution is equivalent to the prior on $X$ so that the event $X = \uvtheta$ is uninformative; indeed, the likelihood is constant when there is no information about $\uvtheta$ and constant likelihoods bring no information about the quantity of interest. If however the coin was not necessarily fair with the probability of $X = 1$ being denoted $p$ and if there was some knowledge about $\uvtheta$ such that $f_{\uvtheta}(0) = \alpha$ and $f_{\uvtheta}(1) = \beta$ with $\max\{\alpha,\beta\} = 1$, then we would have
$$
p_X(\phi \given \uvtheta = X) =  \dfrac{\alpha(1-p)\phi(0) + \beta p \phi(1)}{ \alpha(1-p) + \beta p }.
$$
for any function $\phi$ on $\{0,1\}$. Given the form of \eqref{eq:XGivenE}, it is natural to conclude that this posterior distribution is characterising $X$. Yet, the conclusion reached in \cite{Gelman2006}, which follows from interpreting it as a posterior for $\uvtheta$, is also mathematically correct. This could indeed be seen as paradoxical if $\uvtheta$ was a deterministic uncertain variable. However, in this formulation, $\uvtheta$ is a non-deterministic uncertain variable of the form $\uvtheta(\bscdot, X)$ and can therefore be random as well. In this interpretation, $\uvtheta$ is indeed a deterministic transformation of $X$, which is the identity in this case.

The o.p.m.\ describing the other possible ordering, i.e.\ $\uvtheta$ first and then $X$, is characterised by
$$
\oP_{\uvtheta,X}(\varphi) = \dfrac{1}{2} \max \big\{ \varphi(0,0) + \varphi(0,1), \varphi(1,0) + \varphi(1,1)\big\},
$$
for any function $\varphi$ in $\{0,1\}^2$. This ordering yields similar results but with reversed roles for $X$ and $\uvtheta$, that is
$$
\oP_{\uvtheta}(\phi \given X = \uvtheta) = \dfrac{\max\{ \alpha(1-p)\phi(0), \beta p \phi(1)\}}{\max\{\alpha(1-p),\beta p\}},
$$
for any function $\phi$ on $\{0,1\}$, in general. Once again, considering a fair coin and an uninformative possibility function $f_{\uvtheta}$ yields $\oP_{\uvtheta}(\phi \given X = \uvtheta) = \oP_{\uvtheta}(\phi)$ for any function $\phi$ on $\{0,1\}$. However, in this case, reinterpreting $\oP_{\uvtheta}(\phi \given X = \uvtheta)$ as a posterior o.p.m.\ describing $X$ yields a more challenging situation: the random variable $X$ seems to have lost its randomness through conditioning on $\uvtheta = X$. Yet, this can also be explained by recalling that the value of $\uvtheta$ was fixed before $X$ was sampled; so that the only possible interpretation of $\uvtheta = X$ is ``the \emph{realisation} of $X$ happened to be equal to $\uvtheta$''. The o.p.m.\ $\oP_{\uvtheta}(\phi \given X = \uvtheta)$ is therefore describing the information available about the realisation of $X$. 

\section{Conclusion}

By introducing analogues of some fundamental concepts belonging to the probabilistic and Bayesian paradigms and by showing that important properties of these concepts are preserved when considering o.p.m.s instead of probability measures, we have provided a starting point for the development of statistical methods based on the considered principles. From a modelling viewpoint, the addition of deterministic uncertainty to the standard probabilistic framework brings additional flexibility and allows for expressing very low levels of information which would have resulted in improper prior/posterior distributions in a probabilistic context. The flexibility of the proposed inference framework does not come at the expense of practical limitations; in fact, pragmatic approaches such as the empirical Bayes method and generalised Bayesian inference follow from our o.p.m-based models. By combining possibility theory and probability theory in a principled but intuitive manner, the proposed approach allows for addressing some of the common criticisms of existing theories based on generalisations of the probabilistic paradigm.

Future research will aim at identifying areas of statistics and machine learning where the proposed quantification of uncertainty will be the most relevant. Since possibility theory relies on optimisation rather on integration, the proposed approach is likely to be useful for optimisation under uncertainty such as in reinforcement learning. The use of possibility functions as priors could also help to address some of the challenges in objective Bayes. In addition, the ability to conduct statistical analysis for o.p.m.-based methods could lead to the derivation of further theoretical guarantees for generalised Bayesian inference.

\bibliographystyle{abbrv}
\bibliography{Uncertainty}

\appendix

\section{General pull-back of possibility function}
\label{sec:pullback}

Let $\uvpsi$ and $\uvtheta$ be two deterministic uncertain variables on $\Psi$ and $\Theta$ respectively, described by $f_{\uvpsi}$ and $f_{\uvtheta}$ respectively. The pull-back operation has been defined as follows: if $\uvpsi = \zeta(\uvtheta)$ for some given mapping $\zeta : \Theta \to \Psi$, and if the possibility function $f_{\uvpsi}$ is such that $\sup f_{\uvpsi} \circ \zeta = 1$ then one can deduce a possibility function for $\uvtheta$ as $f_{\uvtheta} = f_{\uvpsi} \circ \zeta$.

The assumption that $\sup f_{\uvpsi} \circ \zeta = 1$ is meaningful since $f_{\uvpsi}$ should actually be supported by the image $\zeta[\Theta]$ of $\Theta$ by $\zeta$ if we know that $\uvpsi = \zeta(\uvtheta)$. If we first learn about $\uvpsi$ and then are given the information that there exist a deterministic uncertain variable $\uvtheta$ and a mapping $\zeta$ such that $\uvpsi = \zeta(\uvtheta)$, then we can try to condition on the event $A = \{\uvpsi = \zeta(\uvtheta)\}$. Since $A$ is an event in $\Omega_{\u}$, we use the possibility function $f$ for the conditioning: for any $\omega_{\u} \in \Omega_{\u}$,
$$
f(\omega_{\u} \given A) = \ind{A}(\omega_{\u})\dfrac{f(\omega_{\u})}{f(A)} = \ind{A}(\omega_{\u})\dfrac{f_{\uvpsi}(\uvpsi(\omega_{\u}))}{\sup_{\omega'_{\u} \in A}f_{\uvpsi}(\uvpsi(\omega'_{\u}))}
$$
The conditional possibility function describing $\uvpsi$ given $A$ can then be computed as 
$$
f_{\uvpsi}(\psi \given A) 
= \sup\bigg\{ \ind{A}(\omega_{\u})\dfrac{f_{\uvpsi}(\uvpsi(\omega_{\u}))}{\sup_{\omega'_{\u} \in A}f_{\uvpsi}(\uvpsi(\omega'_{\u}))} : \omega_{\u} \in \uvpsi^{-1}[\psi] \bigg\},
$$
for any $\psi \in \Psi$, which is implies that
\eqns{
f_{\uvpsi}(\psi \given A) = \ind{\zeta[\Theta]}(\psi) \dfrac{f_{\uvpsi}(\psi)}{ \sup_{\psi' \in \zeta[\Theta]} f_{\uvpsi}(\psi')}, \qquad \psi \in \Psi.
}
We can then define $f_{\uvtheta}$ as the composition between $f_{\uvpsi}(\bscdot \given A)$ and $\zeta$.

\section{Sufficiency of a deterministic uncertain variable}
\label{sec:sufficiency}

Let $\uvtheta$ be a deterministic uncertain variable on $\Theta$ and let $\calX$ be an uncertain variable of the form $Z(\uvtheta,\bscdot)$. The dependency of $\bbP$ on $\omega_{\u}$ implies that an additional formal condition needs to be introduced in order to define the conditional law of $\calX$ given $\uvtheta$. This condition is introduced in the following definition.

\begin{definition}
\label{def:sufficiency}
Let $(\sfE,\calE,P(\bscdot \given s))$ be a probability space for any $s$ in a set $S$, let $\uvtheta : S \to \Theta$ and let $\calX : S \times \sfE \to \sfX$ be such that $\calX(s,\bscdot)$ is measurable for any $s \in S$, then $\uvtheta$ is said to be \emph{sufficiently informative} for $\calX$ w.r.t.\ $\{P(\bscdot \given s)\}_{s \in S}$ if, for all $\theta \in \Theta$, it holds that
\eqnl{eq:sufficiency}{
\calX(\omega_{\u}, \bscdot)_*P(\bscdot \given \omega_{\u}) = \calX(\omega_{\u}', \bscdot)_*P(\bscdot \given \omega'_{\u})
}
for any $\omega_{\u}, \omega'_{\u} \in \uvtheta^{-1}[\theta]$.
\end{definition}

The notation $\calX(\omega_{\u}, \bscdot)_*P(\bscdot \given \omega_{\u})$ refers to the probability measure induced on $\sfX$ by the pushforward of the probability measure $P(\bscdot \given \omega_{\u})$ by the random variable $\calX(\omega_{\u}, \bscdot)$. The name of the property introduced in Definition~\ref{def:sufficiency} is meant to be related to the concept of sufficiency for statistics since the two notions are related: $\uvtheta$ is sufficiently informative for $\calX$ if the conditional law of $\calX$ given $\uvtheta$ is well defined without the need to mention the parameter $s \in S$. 
Property \eqref{eq:sufficiency} can be seen as an analogue of a measurability condition for the function $\omega_{\u} \mapsto \calX(\omega_{\u}, \bscdot)_*P(B \given \omega_{\u})$ for any measurable subset $B$ of $\sfX$. This property is automatically verified for any $\theta \in \Theta$ that is not in the co-domain of $\uvtheta$ since $\uvtheta^{-1}[\theta]$ is empty in this case.

When $\sfE = \Omega_{\r}$ and $S = \Omega_{\u}$, the deterministic uncertain variable $\uvtheta$ is sufficiently informative for the uncertain variable $\calX$ w.r.t.\ $\{\bbP(\bscdot \given \omega_{\u})\}_{\omega_{\u} \in \Omega_{\u}}$ if the conditional law of $\calX$ given $\uvtheta = \theta$ is well defined. Indeed, in this case we can introduce the probability measure $p_{\calX}(\bscdot \given \theta)$ on $\sfX$ parametrised by $\theta \in \Theta$ through
\eqns{
p_{\calX}(B \given \theta) = \bbP\big( \calX(\omega_{\u}, \bscdot) \in B \given \omega_{\u} \big)
}
for all measurable subset $B$ of $\sfX$ and for any $\omega_{\u}$ such that $\uvtheta(\omega_{\u}) = \theta$. It would be too restrictive to assume that $\bbP(\bscdot \given \omega_{\u})$ is fully specified by $\uvtheta$ since this would imply that $\uvtheta$ characterises all random phenomena in $\Omega_{\r}$ instead of just the ones appearing in $\calX$. Conversely, it is not sufficient for an uncertain variable $\calX$ to be of the form $Z(\uvtheta,\bscdot)$ to ensure that it is sufficiently informative since $\bbP(\bscdot\given \omega_{\u})$ could vary within elements of the partition of $\Omega_{\u}$ induced by $\uvtheta$; yet, \eqref{eq:sufficiency} can be simplified in this case to: for all $\theta \in \Theta$, it holds that
\eqns{
Z(\theta, \bscdot)_*\bbP(\bscdot \given \omega_{\u}) = Z(\theta, \bscdot)_*\bbP(\bscdot \given \omega'_{\u}), \qquad \omega_{\u}, \omega'_{\u} \in \uvtheta^{-1}[\theta]
}
or, alternatively, that $X(\theta, \bscdot)_*\bbP(B \given \bscdot)$ is constant over $\uvtheta^{-1}[\theta]$ for any measurable subset $B$ of $\sfX$.

\end{document}